\documentclass[review]{elsarticle}

\usepackage{hyperref}\usepackage{amsmath}

\journal{Modern Physics Letters A}

\begin{document}

\begin{frontmatter}

\title{The Atmosphere of a Black Hole}

\author{Metin Arik}
\ead{metin.arik@boun.edu.tr}
\author{Yorgo Senikoglu}
\ead{yorgo.senikoglu@boun.edu.tr}
\address{Department of Physics, Bo\u{g}azi\c{c}i University, Bebek, Istanbul, Turkey}


\begin{abstract}
We present a static, isotropic fluid solution around a black hole and its effects on the field equations and the horizon. We offer the solutions, their descriptions and we comment on their shortcomings. We derive from the proposed metric the density, the pressure, the equation of state, the temperature near and far of the black hole and the equation of state parameter. We show that for a special case, when $r\rightarrow\ 0$ the solution behaves as $p=\frac{\rho}{3}$, the equation of state for radiation and for $r\rightarrow\infty$ we observe that $p=0$, the equation of state for dust.
\end{abstract}

\begin{keyword}
\texttt{Physics of Black Holes, General Formalism, Exact Solutions}
\end{keyword}

\end{frontmatter}

\section{Introduction}
In Einstein's theory of general relativity, static solutions with spherically symmetric spacetimes could be used to describe the relativistic spheres in astrophysics. This is why different techniques and analysis are worked through to attain exact solutions. Durgal \textit{et al}., \cite{A} modeled the physical requirements for a neutron star. There are models \cite{B} \cite{C} for static, spherically symmetric solutions which provide leading paths. 
John \textit{et al}., \cite{D} derived an exact isotropic solution which reduces to a recurrence equation with variable, rational coefficients of order three. 
\newpage
\noindent
Semiz \cite{E} searched for spherically symmetric solutions for perfect fluids to include the possibilities of dark energy and phantom energy pervading the spacetime.   
Magli \cite{F} proposed simple models of 'black hole interiors' which satisfy the weak energy condition and the matter content is specified by an equation of state of the elastic type. Dotti \textit{et al}., \cite{G} studied spherically symmetric solutions that have a regular horizon and satisfy the weak and dominant energy conditions outside the horizon.
Emparan \textit{et al}., \cite{H} examined spaces with some rotational symmetries which provided a rich spectrum of different phases of black objects with distinct topologies for horizons. Pons \textit{et al}., \cite{I} considered Einstein-Gauss-Bonnet black holes, a new aspect is that there can occur a non-central naked singularity, that can be averted by imposing a range for black hole mass. Gannouji \textit{et al}., \cite{J} inspected the existence and stability of static black holes in Lovelock theories and derived the equation of stability from action.
Davidson \textit{et al}., \cite{K} have proven that any static metric with a Killing horizon in the presence of a perfect fluid, is necessarily a Schwarzschild solution with a vanishing proper energy density and a vanishing proper pressure. Solutions describing black holes embedded in gradually increasing perfect fluid and the phenomenon for $p=\omega\rho$ were demonstrated. Leading to the proof that the only perfect fluid $p=\omega\rho$ static, spherically symmetric black hole solution is the Schwarzschild solution with vanishing p and $\rho$.

The motivation that drove us to this paper was the possibility if a solution satisfying only the isotropy condition existed in a static spacetime. In the following section we will propose a two parameter solution of the Field Equations that lead to interesting findings. One of the parameters is related to the mass of the black hole and the other to the energy density in the universe.
\newpage

\section{Solution}

Let us write the metric in a static spacetime with a function f(r).
\begin{multline}
ds^2=\frac{(1-f(r))^2}{(1+f(r))^2}dt^2-(1+f(r))^4 (dr^2+r^2 d{\Omega}^2)\\where \hspace{2mm}d{\Omega}^2 = d{\theta}^2+\sin^2\theta d{\phi}^2.
\end{multline}
\noindent
The motivation for this ansatz was to investigate the singularity at $r=0$. Instead of putting a solution like $\frac{1}{r}$ directly we searched for a more general form for a solution.
The calculations lead us to the following non-vanishing components of the Einstein tensor in the orthonormal basis
${G}_{tt}$, ${G}_{rr}$ and ${G}_{\theta\theta}={G}_{\phi\phi}$.

We will not bother the reader with the long and exhausting details of these components at this point.

The only condition that we impose is the following, in the orthonormal basis:
\begin{equation}
{G}_{rr}={G}_{\theta\theta}={G}_{\phi\phi}.
\end{equation}

What we obtain is rather intriguing, the unique solution that satisfies (2) is:
\begin{equation}
f(r)=\frac{1}{\sqrt{a^2r^2+2b}}=(a^2r^2+2b)^{-1/2}. 
\end{equation}
where a and b are constants.

\noindent
We obtain the following non-vanishing components of the Einstein tensor in the orthonormal basis:
\begin{equation}
{G}_{tt}=\frac{24a^2b}{(\sqrt{a^2r^2+2b}+1)^5},
\end{equation}
\begin{multline}
{G}_{rr}={G}_{\theta\theta}={G}_{\phi\phi}=\frac{8a^2b}{(\sqrt{a^2r^2+2b}-1)(\sqrt{a^2r^2+2b}+1)^5}.
\end{multline}

\noindent
For further purposes we denote the equation of state parameter:
\begin{equation}
\nu=\frac{1}{3(\sqrt{a^2r^2+2b}-1)}=\frac{1}{3(\frac{1}{f(r)}-1)}
\end{equation}
${G}_{rr}={G}_{\theta\theta}={G}_{\phi\phi}=p$ and ${G}_{tt}=\rho$.
where p and $\rho$ are respectively the pressure and density.

\noindent
i.e. with (3), $a=\frac{2}{GM}$ and $b=\frac{8}{3}G^3M^2\rho_{H}$
\begin{equation}
\rho=\frac{32\rho_{H}}{(\frac{1}{f(r)}+1)^5} \hspace{2mm}and \hspace{2mm} p=\nu\rho
\end{equation}
$\rho_{H}$ being the energy density at the horizon.

\noindent
The non-vanishing components of the Weyl Tensor are:
\begin{equation}
C_{0101}=-2C_{0202}=-2C_{0303}=2C_{1313}=2C_{2323}=\frac{4a^4r^2}{f(r)(\frac{1}{f(r)}+1)^6}
\end{equation}

\section{Discussions}
\paragraph{Singularity}

We immediately observe the horizon singularity at $f(r)=1$, we note that the pressure $p\rightarrow\infty$ whereas the density and the Weyl Tensor components are finite at the horizon. 
On the other hand for $r\rightarrow\infty$, we see that the equation of state paramter $\nu$ goes to zero and the components of the Weyl tensor signal a massive object of mass M. Far from the black hole, matter is approximated as stationary dust particles which produce no pressure.

\paragraph{Can the horizon be circumvented?}

\noindent
The answer to this question can be given by a choice of b in f(r).
\noindent
We can circumvent this singularity for $\sqrt{a^2r^2+2b}>1$ which gives us for all r , $b>\frac{1}{2}$.
\vspace{4mm}

\noindent
Let us consider and analyze the situation where we have to choose b so that the positivity of the energy condition is not violated.

\noindent
Noting $p=\nu\rho$ with $\nu=\frac{1}{3(\sqrt{a^2r^2+2b}-1)}$, as $r\rightarrow 0$,
we calculate b for $\nu\leq 1$ i.e. $p\leq\rho$.

\noindent
We simply obtain by doing the calculation that $b\geq\frac{8}{9}$.

\vspace{4mm}
\noindent
Next we calculate b so that $p=\frac{\rho}{3}$, as $r\rightarrow 0$ and we get $b=2$.
That is to say that if we choose b=2, we have as $r\rightarrow 0$,
$p=\frac{\rho}{3}$ which is the radiation equation of state and as $r\rightarrow\infty$,
$p=0$ the dust equation of state. And the singularity is circumvented.

\paragraph{Temperature}
Another interesting aspect of this metric with f(r), the density and the pressure formulations that we have derived is that we can calculate via 
\begin{equation}
\frac{d}{dT}(\frac{\rho+p}{T})=\frac{1}{T}\frac{d\rho}{dT}
\end{equation} the temperature as a function of r.

\noindent
Doing the necessary calculations we have:
\begin{equation}
\frac{T}{T_{\infty}}=\frac{1+f(r)}{1-f(r)}
\end{equation}
where $T_{\infty}$ is the temperature at $r\rightarrow\infty$.
We see that the change of temperature is quite mild unless there is a horizon in which case the temperature is infinite at the horizon. We note that the equation of state parameter is:
\begin{equation}
\nu=\frac{1}{6}(\frac{T}{T_{\infty}}-1).
\end{equation}
and observe that $\nu$ has a linear dependence on temperature.

\vspace{3mm}
\section{Conclusion}
This solution suggests that the vacuum black hole solution attains a singular horizon even if a small amount of isotropic matter-energy is introduced. In this case one expects a non static solution with isotropic matter-energy falling into the black hole.

\noindent
In case of a large amount of isotropic matter-energy, the horizon disappears. We physically expect that again, matter falling into the black hole eventually causes a singular horizon to appear, with the horizon getting bigger as matter-energy keeps falling into the black hole. Thus investigation of such non-static solutions is relevant.

\noindent
The fact, $\nu>1$, near the singular horizon is probably due to the fact that interactions other than gravity are neglected. It is a well know fact that the increase of pressure as real matter falls into the black hole causes thermonuclear fusion to occur.

From cosmological solutions, we know that matter-energy in a spatially flat universe causes the universe to expand. This may be relevant for our solution which is static and the tendancy of matter-energy to expand is balanced by the pull of the black hole.

\noindent
For large values of the dimensionless parameter b, the equation of state remains close to the dust equation of state for all r. For values of $b>\frac{1}{2}$, there is no horizon and the maximum value of energy density $\rho_{0}$ is attained for r=0.

\noindent
For $b>>\frac{1}{2}$, the relation between b and $\rho_{0}$ is
\begin{equation}
b=(\frac{9}{2\pi^2G^6M^4\rho_{0}^2})^\frac{1}{3}.
\end{equation}
Hence the energy density of the universe prevents small mass objects from ever attaining a horizon.

\bibliography{mybibfile}

\begin{thebibliography}{10}
\expandafter\ifx\csname url\endcsname\relax
  \def\url#1{\texttt{#1}}\fi
\expandafter\ifx\csname urlprefix\endcsname\relax\def\urlprefix{URL }\fi
\expandafter\ifx\csname href\endcsname\relax
  \def\href#1#2{#2} \def\path#1{#1}\fi

\bibitem{A}
M.C.Durgapal, R.Bannerji, New analytical stellar model in general relativity,
  Phys.Rev.D 27 (1983) 328.

\bibitem{B}
H.Stephani, D.Kramer, M.A.H.Mac{C}allum, C.Hoeselaers, E.Herlt, Exact
  {S}olutions of {E}instein's {F}ield {E}quations, Cambridge University Press,
  2003.

\bibitem{C}
M.S.R.Delgaty, K.Lake, Physical {A}cceptability of {I}solated, {S}tatic,
  {S}pherically {S}ymmetric, {P}erfect {F}luid {S}olutions of {E}instein's
  {E}equations, Comput.Phys.Commun. 115 (1998) 395--415.

\bibitem{D}
A.J.John, S.D.Maharaj, An exact isotropic solution, Nuovo Cimento B121 (2006)
  27--33.

\bibitem{E}
I.Semiz, All 'static' spherically symmetric perfect fluid solutions of
  {E}instein's equations with constant equation of state parameter and finite
  polynomial 'mass function', Rev.Math.Phys. 23 (2011) 865--882.

\bibitem{F}
G.Magli, Physically valid black hole interior models, Rep.Math.Phys 44 (1999)
  407--412.

\bibitem{G}
G.Dotti, R.J.Gleiser, Static black hole solutions with a self interacting
  conformally coupled scalar field, Phys.Rev.D 77 (2008) 104035.

\bibitem{H}
R.Emparan, H.S.Reall, Black holes in {H}igher {D}imensions, Living.Rev.Rel. 11
  (2008) 6.

\bibitem{I}
J.M.Pons, N.Dadhich, On static black holes solutions in {E}instein and
  {E}instein-{G}auss-{B}onnet gravity with topology {S}{O}(n)x{S}{O}(n),
  arXiv'gr-gc' 1408.6754.

\bibitem{J}
R.Gannouji, N.Dadhich, Stability and existence analysis of static black holes
  in pure {L}ovelock theories, Class.Quantum.Grav. 31 (2014) 165016.

\bibitem{K}
A.Davidson, S.Rubin, Can an evolving universe host a static event horizon,
  Phys.Rev.D 86 (2012) 104061.

\end{thebibliography}

\end{document}